\documentclass{iopconfser}
\usepackage{braket}
\usepackage{amsmath}
\usepackage{graphicx}
\usepackage{import}
\usepackage[absolute,overlay]{textpos}
\usepackage{lipsum}
\usepackage{multirow}
\usepackage{makecell}
\usepackage[dvipsnames]{xcolor}

\definecolor{mediumpersianblue}{rgb}{0.0, 0.4, 0.65}
\definecolor{persianred}{rgb}{0.8, 0.2, 0.2}

\newcommand{\Tr}{\text{Tr}}

\newcommand{\HSnorm}[1]{\lVert#1\rVert_{\text{HS}}}
\newcommand{\HSprod}[2]{\langle#1,#2\rangle_{\text{HS}}}
\newcommand{\pardiff}[2]{\frac{\partial #1}{\partial #2}}
\newcommand{\DBR}[2]{e^{s#1}#2e^{-s#1}}
\newcommand{\diag}{\text{diag}}
\newcommand{\W}{\hat W}
\newcommand{\D}{\hat D}
\newcommand{\Hhat}{\hat H}
\newcommand{\h}{\hat H}
\newcommand{\J}{\hat J}
\def\v{\hat V}

\newcommand{\Z}{\hat Z}
\newcommand{\X}{\hat X}
\newcommand{\Y}{\hat Y}
\newcommand{\A}{\hat A}

\begin{document}

\begin{textblock*}{4cm}(15cm,2cm)
    TIF-UNIMI-2024-8
\end{textblock*}

\title{Strategies for optimizing double-bracket quantum algorithms}

\author{Li Xiaoyue$^{1}$,  Matteo Robbiati$^{2,3}$,  Andrea Pasquale$^{3,4,5}$, Edoardo Pedicillo$^{3,4}$,
Andrew Wright$^{6}$, Stefano Carrazza$^{2,3,4,5}$ and  Marek Gluza$^{1}$ }

\affil{$^1$School of Physical and Mathematical Sciences, Nanyang Technological
University, 21 Nanyang Link, 637371 Singapore, Republic of Singapore}
\affil{$^2$European Organization for Nuclear Research (CERN), Geneva 1211, Switzerland}
\affil{$^3$TIF Lab, Dipartimento di Fisica, Universit\`a degli Studi diMilano}
\affil{$^4$Quantum Research Centre, Technology Innovation Institute, Abu Dhabi, UAE}
\affil{$^5$INFN Sezione di Milano, Milan, Italy}
\affil{$^6$Institute of Physics, Ecole Polytechnique Fédérale de
Lausanne (EPFL), Lausanne, Switzerland.}

\email{marekludwik.gluza@ntu.edu.sg}

\begin{abstract}
Recently double-bracket quantum algorithms have been proposed as a way to
compile circuits for approximating eigenstates.
Physically, they consist of appropriately composing evolutions under an input Hamiltonian
together with diagonal evolutions.
Here, we present strategies to optimize the choice of the double-bracket evolutions
to enhance the diagonalization efficiency. This can be done by finding optimal
generators and durations of the evolutions. We present numerical results regarding
the preparation of double-bracket iterations, both in ideal cases where the algorithm's
setup provides analytical convergence guarantees and in more heuristic cases,
where we use an adaptive and variational approach to optimize the generators of the evolutions.
As an example, we discuss the efficacy of these optimization strategies when considering
a spin-chain Hamiltonian as the target.
To propose algorithms that can be executed starting today, fully aware
of the limitations of the quantum technologies at our disposal, we finally present
a selection of diagonal evolution parametrizations that can be directly compiled into
CNOTs and single-qubit rotation gates. We discuss the advantages and limitations of
this compilation and propose a way to take advantage of this approach
when used in synergy with other existing methods.
\end{abstract}

\section{Introduction}

Preparing the eigenstates of a target physical system is crucial in fields
like physics, chemistry, and combinatorial optimization. Various strategies exist
within both classical~\cite{PhysRevLett.69.2863, Carleo_2017, VariationalMC,BiamonteTN}
and quantum computing domains. In the quantum realm, high-precision algorithms like
quantum phase estimation~\cite{PRXQuantum.3.040305, ge2018fastergroundstatepreparation}
(QPE) await technological advancements for practical use,
while near-term routines such as variational quantum eigensolvers~\cite{Peruzzo_2014}
(VQEs) are more feasible. This work focuses on a family of algorithms suitable
for near-term quantum devices: double-bracket quantum algorithms (DBQAs).

These algorithms have been proposed to prepare eigenstates through a recursive double-bracket
iteration (DBI) strategy~\cite{Gluza_2024} and they can be compiled into quantum circuits.
In fact, originally, Ref.~\cite{Gluza_2024} reduced DBQA compilation from the
abstract DBI equations to a sequence of queries to Hamiltonian simulation which can
be compiled by known methods~\cite{Low2019hamiltonian,RandomCompiler,HamiltonianSimulationTruncatedSeries,ChildsSu},
but left open quantitative questions about the precise cost of compiling DBIs into circuits.
In particular, the depth of DBQA circuits grows exponentially with the number of recursion steps.

Despite this, recent studies~\cite{boostvqe} show that when initialized properly, DBQAs
can substantially improve the eigenstate fidelity reached by quantum algorithms such as VQEs in just one or two steps.

This trade-off between computational cost and gain is crucial: DBQAs must be used
synergistically with other approximation techniques and double-bracket iterations must be optimized
to provide maximum benefit in a few recursion steps.

Moreover, DBQAs can be optimized to maximize diagonalization gain in every step.
Specifically, we adopt optimization methods to
accelerate the preparation of a target eigenstate.
In this proceedings, we aim to study in depth such optimization techniques.
We present numerical results implemented in Qibo~\cite{Efthymiou_2021, Carrazza_2023, Efthymiou_2022},
as used in Ref~\cite{boostvqe}.

\section{Double-bracket quantum algorithms for diagonalization}
\label{secDBQA}

The building blocks of the double-bracket quantum algorithms are double-bracket rotations (DBRs),
which iteratively update a target Hamiltonian $\h_0$ following

\begin{align}
    \Hhat_{k+1}=e^{s\W_k}\Hhat_ke^{-s\W_k} \equiv \hat{R}_k^{\dagger} \Hhat_k \hat{R}_k\ .
    \label{eq:dbr0}
\end{align}
Here $\W_k=[\D_k, \Hhat_k]$ is a commutator involving $\D_k$ which is a diagonal hermitian operator
and $\hat{R}_k=e^{-s\W_k}$ is the unitary operator implementing the DBR.
In order to see why DBRs have diagonalizing properties, we motivate the above ansatz $\W_k=[\D_k, \Hhat_k]$, by letting $\W_k$ be any anti-hermitian operator.
Let $\Delta(A)$ be the diagonal restriction of matrix $A$ and $\sigma(A)$ the off-diagonal restriction, s.t. $A=\Delta(A)+\sigma(A)$.
Performing a Taylor expansion of  of $\sigma(\Hhat_{k+1})$ in Eq.~\eqref{eq:dbr0} around $s=0$ and using that
$\pardiff{}{s}||\sigma(\DBR{\W_k}{\Hhat_k})||^2_{\text{HS}, s=0} =-2\HSprod{\W_k}{[\Delta(\Hhat_k), \Hhat_k]}$
we get
\begin{align}
    \sigma(\Hhat_{k+1})  =\sigma(\Hhat_{k})  -2s\HSprod{\W_k}{[\Delta(\Hhat_k), \Hhat_k]} + \ldots\ ,
\label{eq:sigma_decrease}
\end{align}
which implies that the rotation reduces the off-diagonal norm of the Hamiltonian $\Hhat_{k+1}$
compared to $\Hhat_{k}$ if
\begin{align}
\label{eq:CS_angle}
    {\HSprod{\W_k}{[\Delta(\Hhat_k), \Hhat_k]}}>0\ .
\end{align}

The choice, $\W_k = [\Delta(\Hhat_k), \Hhat_k]$, emerges naturally and guarantees an
initial drop in the off-diagonal norm and is known as the canonical bracket.
Additionally, if we choose $\W_k=[\D_k, \Hhat_k]$, where $\D_k$ is a diagonal hermitian operator which may be 
easier to implement than $\Delta(\Hhat_k)$,
the overlap is likely to induce a $\sigma$-decrease in Eq.~\eqref{eq:sigma_decrease}.
Therefore, this ansatz is largely flexible in devising diagonalization DBIs outside of using the canonical bracket.

\subsection{Types of DBIs}
In this work, we will consider three types of DBIs:
the first two are derived from double-bracket flow formulations
(the continuous version of the one analyzed here), while the third is a
variational approach to the problem of optimizing the generators $\D_k$ of the double-bracket
evolutions.

The first DBI formulation is inspired by Brokett's work~\cite{BROCKETT199179}, which introduced
a double-bracket flow involving a fixed diagonal operator $\hat{N}$. Afterwards,
Moore, Mahony, and Helmke~\cite{moore_mahony_helmke} have considered double-bracket iterations where each
double-bracket rotation uses $\hat{N}$ as diagonal operator: $\D_k=\hat{N}$.
They concluded that if $\hat N$ has a non-degenerate spectrum then BHMM converges to a diagonal fixed-point $\Hhat_\infty$
and the only stable fixed-point is the one where $\Hhat_\infty$ has the same sorting of eigenvalues as $\hat{N}$.
Hence, we call this strategy the Brokett-Helmke-Moore-Mahony (BHMM) DBI.

The second double-bracket flow is based on works by Głazek, Wilson, and Wegner~\cite{PhysRevD.48.5863,flow_equation} (GWW).
The corresponding GWW DBI arises recursively by setting $\D_k = \Delta(\Hhat_k)$, which makes $\W_k$ the canonical bracket.
This choice results in an unconditional $\sigma$-decrease where Eq.~\eqref{eq:sigma_decrease} becomes
\begin{align}
    \pardiff{}{s}||\sigma(\DBR{\W_k}{\Hhat_k})||^2_{\rm{HS}, s=0}=-2||\W_k||_{\rm HS}^2\ .
\label{eq:sigma_decrease2}
\end{align}

The third DBI paves the way for a variational approach. Given each $\Hhat_k,$
one can consider introducing a parametric form of $\D_k$ and optimizing its parameters
to maximize the decrease of a cost function which quantifies the effectiveness of diagonalization. One possible choice for the cost function
could be the off-diagonal norm of the target Hamiltonian. In Sec~\ref{sec:cost_functions}
we introduce some other possible cost functions.
We will call adaptive such strategies where we optimize $\D_k$ for each $\Hhat_k$.
They have the advantage of possibly being more efficient but, due to implicit numerical optimization,
their convergence analysis is precluded as opposed to GWW and BHMM DBIs.

\subsection{\label{sec:cost_functions} Cost functions for diagonalization}
When defining an optimization procedure to find a good candidate as a double-bracket
rotation generator, it is crucial to select a proper cost function whose value has
to be optimized.
We next review possible cost functions and 
comment on their role in a prospective quantum computation. 

A first possible choice ($f_1$ in Tab.~\ref{tab:costs}) is the off-diagonal norm, i.e. the Hilbert-Schmidt
norm of the restriction to the off-diagonal of an input,
which is motivated by the monotonicity relation~\eqref{eq:sigma_decrease}. 

A second cost function ($f_2$ in Tab.~\ref{tab:costs}) is inspired by the work of Brockett~\cite{BROCKETT199179},
who originally considered least-square functional optimization $\|\hat A- \hat N\|_{\rm HS}$
where $\hat A$ (in our case $\hat A = \hat H_0$ but the setting is more general) 
was being varied while $\hat N$ was a fixed target matrix which can be chosen to be diagonal
 without loss of generality.
If the DBI considered is BHMM then we can set $\D_k = \hat N$ and indeed
$f_2(\Hhat_k) = \|\Hhat_k- \hat N\|^2 - \|\Hhat_0\|^2$,
agreeing with Brockett's cost function because $\|\Hhat_k\|$ is constant due to unitary invariance.
As reported in $f_2$ formula from Tab.~\ref{tab:costs}, the norm $||\D_k||_{\rm HS}^2$ is needed to fix normalization.
For the GWW DBI we have $\D_k=\Delta(\Hhat_k)$, the $f_2$ cost function reduces to  $ f_1$
because $\Hhat_k = \Delta(\Hhat_k) +\sigma(\Hhat_k)$.

A third choice is to consider the energy expectation for a given state ($f_3$ in Tab.~\ref{tab:costs}).
Minimizing this value aims to find the ground energy and state of the target Hamiltonian,
making it a good cost function to utilize in conjunction with the VQE~\cite{boostvqe}.

Finally, we can use the energy fluctuation of a reference state $\ket{\psi}$ ($f_4$ in Tab.~\ref{tab:costs}). 
It allows to detect whether $\ket \psi$ is converging to an eigenstate of $\Hhat$.
 A few steps of DBI can usually produce a
good eigenstate approximation for some states $\ket{\psi}$, making the energy
fluctuation an appropriate measure of eigenstate approximation which is necessary for full diagonalization.

\renewcommand{\arraystretch}{2}
\begin{table*}[ht]
\centering
\small
\begin{tabular}{ccccc}
\hline \hline
\textbf{Function} & Off-diagonal norm ($f_1$) & Least-squares ($f_2$) & Energy ($f_3$) & Energy fluctuation ($f_4$)\\
\hline
\textbf{Formula} & $||\sigma(\Hhat_k)||_{\rm HS}$ & $\frac{1}{2}||\D_k||_{\rm HS}^2-\Tr(\Hhat_k\D_k)$ & $\langle \psi | \hat{H}| \psi \rangle$ & $\sqrt{(\langle\psi|\Hhat_k^2|\psi\rangle-\langle\psi|\Hhat_k|\psi\rangle^2)}$ \\
\hline
\textbf{Utility} & Diagonalization & Diagonalization & \makecell{Ground state \\ preparation} & \makecell{Eigenstate \\ preparation} \\
\hline \hline
\end{tabular}
\caption{\label{tab:costs} Cost functions DBQAs optimization with possible applications.}
\end{table*}

As shown in Fig.~\ref{fig:polynomial_fit}, when we optimize DBRs based on the energy fluctuation of a state $|\psi\rangle$,
while the energy fluctuation is indeed suppressed and the eigenstate is well approximated, the off-diagonal norm is not necessarily reduced.

\subsection{Optimization tasks}
Once a specific DBI is chosen and we decide
to formalize the diagonalization by minimizing a specific cost function $f_*$,
the optimization problem can be divided, in practice, into three tasks:
the \textit{scheduling}, the \textit{generator selection} and the
\textit{compilation}.

The first task involves finding the DBR duration $s_k$ which solves
\begin{equation}
    \text{argmin}_{s_k<s_\text{max}} \bigl[ f_*(\DBR{[\D_k,\Hhat_k]}{\Hhat_k}) \bigr]
    \label{eq:scheduling_task}
\end{equation}
given a diagonalization operator $\D_k$. In Eq.~\eqref{eq:scheduling_task} we limit
the optimization to some $s_\text{max}$ because $s_k$ increases the gate cost of DBI,
see Sec.~\ref{sec:scheduling}.

The \textit{generator selection} task seeks to optimize the diagonal
operator $\D_k$ in the rotation generator $\W_k$. This entails parametrization
of $\D_k$ appropriate for quantum computing and assumes oracle access to a solver of the
scheduling task.
We quantify the output of the generator selection task by evaluating the diagonalization gain
$f_*(\DBR{[\D_k,\Hhat_k]}{\Hhat_k})$ as detailed in Sec.~\ref{sec:generator}.

Finally, DBR \textit{compilation} means the DBR unitaries $\hat{R}_k$
must be approximated by a circuit made of primitive gates and is discussed in Sec.~\ref{sec:gc}.

\subsection{Models considered}
As practical study cases, we consider two spin-chain Hamiltonians as targets. Firstly,
we tackle the transverse field Ising model (TFIM) with periodic boundary conditions (PBC)
\begin{align}
    \Hhat_{\text{TFIM}} = \sum _{j=1}^L  \X_j \X_{j + 1} + h \Z_j \ .
    \label{eq:TFIM}
\end{align}
Sometimes TFIM is defined as $\sum _{j=1}^L  \Z_j \Z_{j + 1} + h \X_j$ but having diagonalization in mind
we take the 2-qubit interaction to be off-diagonal.
Secondly, we consider the Heisenberg XXZ model again with PBC
\begin{align}
    \Hhat_{\text{XXZ}} = \sum _{j=1}^L (\X_j\X_{j + 1} + \Y_j\Y_{j + 1} + \delta \Z_j\Z_{j + 1} ) \ .
    \label{eq:XXZ}
\end{align}
we will use TFIM and XXZ to demonstrate our observations about optimizing DBI in Qibo.

\section{Scheduling}

\label{sec:scheduling}

\begin{figure}[ht]
    \centering
    \includegraphics[width=1\textwidth, trim = 0 0.4cm 3cm 0, clip]{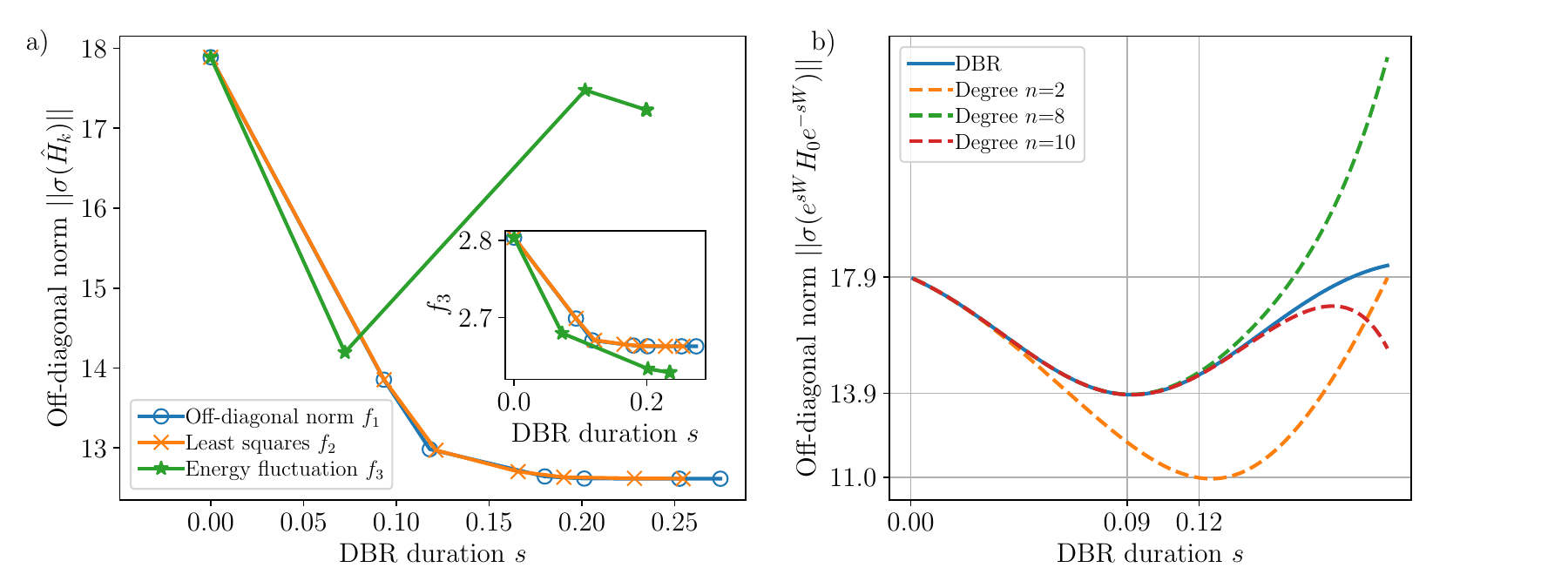}
    \caption{TFIM with $L=5$, transverse field $h=3$ and $\D_0=\Delta(\Hhat_0)$.
    a) We show $f_1$ for DBI with steps optimized using $f_1, f_2$ and $f_3$.
     The inset shows that optimization with $f_3$ is sensitive to the reference state.
    b) We show the validity of the Taylor expansion for various orders $n$.
    For all $n$ the DBR is well approximated for short DBR durations $s<0.04$.
    In order to match the first local minimum of the $\sigma$-decrease curve, we find that a relatively high polynomial degree is needed.
    }
    \label{fig:polynomial_fit}
\end{figure}

The simplest approach to find the optimal DBR duration $s_k$ at each step consists of using a standard
hyper-optimization algorithm. Grid search, random search~\cite{liashchynskyi2019gridsearchrandomsearch},
Tree of Parzen estimators~\cite{tpe}
techniques are all viable for this task and they are available in Qibo.
The scheduling cost functions from Sec.~\ref{sec:cost_functions}
have analytical properties which we can take into account by using Taylor expansion.
For this we define
\begin{equation}
    \Gamma_n = [\W, \Gamma_{n-1}]
\end{equation}
with $\Gamma_0=\hat H$ where $\hat H$ is any operator, e.g. $\hat H_0$ or $\hat H_k$.
The Baker–Campbell–Hausdorff (BCH) formula gives $\DBR{\W}{\Hhat}=\Gamma_0+s\Gamma_1+\frac{s^2}{2}\Gamma_2+O(s^3)$.
Taking the off-diagonal norm as an example, Eq.~\eqref{eq:sigma_decrease} taken up to degree 2
yields a parabola of variable $s$:
\begin{align}
\begin{split}
\label{eq:polynomial_expansion}
    \pardiff{}{s}\HSnorm{\sigma(\DBR{\W}{\Hhat})}^2
    & = s^2 \operatorname{Tr} \left(3 \sigma (\Gamma_1)\sigma (\Gamma_2) + \sigma (\Gamma_0) \sigma (\Gamma_3)\right)
    + 2s \operatorname{Tr} \left( \sigma (\Gamma_1) \sigma (\Gamma_1) + \sigma (\Gamma_0) \sigma (\Gamma_2) \right) \\
    & \quad + 2 \operatorname{Tr} \left( \sigma (\Gamma_0) \sigma (\Gamma_1) \right) + O(s^3)\ .
\end{split}
\end{align}
Subsequently, we can integrate Eq.~\eqref{eq:polynomial_expansion} and obtain the
expansion for the off-diagonal norm $||\sigma(\DBR{\W}{\Hhat})||$ as a polynomial.
Fig.~\ref{fig:polynomial_fit} shows the results of scheduling based on finding minima of Taylor polynomials and
compares the Taylor approximation to the exact $\sigma$-decrease
curve.
More specifically, we find the roots of the Taylor polynomial of Eq.~\eqref{eq:polynomial_expansion} and output the minimizer
 $s$ at the first extremum which by the monotonicity relation will be a minimum, provided that DBR is directed to diagonalization.
This avoids evaluating multiple exponentials for cost function queries and the obtained approximation
can be improved by taking higher-order Taylor polynomials.
Moreover, while the first minimum may not be the global minimum, we have seen cases where
taking the first local minimum outperforms taking the global minimum of the DBR after several iterations.
This indicates that greedy strategies at every step are not necessarily optimal.

\section{Generator selection}

\label{sec:generator}

In this section, we introduce some possible strategies to parametrize diagonal operators.
We begin with ansatze that are motivated analytically and then discuss those suitable for
quantum compiling. Finally, we present numerical results considering a 5-qubit Heisenberg
model as the target and the off-diagonal norm as the cost function.

\subsection{Min-max inspired parametrizations}
Moore, Mahony, and Helmke proved that if $D^*$ is non-degenerate then the BHMM DBI converges~\cite{moore_mahony_helmke}.
A simple example of such an operator is the \textit{min-max parametrization} where the values on the diagonal
are arranged from the smallest to the largest element of the diagonal entries of $\Hhat$ and are equidistant.
In other words, we set $\delta=\frac{d_n-d_1}{n}$ so that
\begin{align}
    \D^*_{\rm min-max}=\sum_{i=1}^{2^L} (\min(\Delta(\Hhat))+i\delta)\ket i\bra i \ .
    \label{eq:minmax}
\end{align}
This ansatz can be difficult to implement on a quantum computer using local gates but  it is an instructive toy model
regarding convergence properties of DBIs, e.g.
it is sufficient if $\D^*$ is non-degenerate.

Since the BHMM DBIs have sorting properties (the fixed point is sorted according to $\D^*$),
we also investigated the inverted \textit{max-min parametrization} with entries like min-max but in the reversed order, i.e.
we use $\max$ instead of $\min$ in Eq.~\eqref{eq:minmax}.
Remarkably, depending on the input basis of $\Hhat$ one of these two orderings might
be closer to the initialization.

To explore the impact of the ordering, we apply a random permutation
to the spectrum of the min-max parametrization. We refer to this strategy as
\textit{shuffled min-max parametrization}.
There are many possible orderings of a min-max parametrization and a random
sample of an ordering can reveal insights that structured min-max and max-min
parametrization does not exhibit.

To further investigate how degeneracy affects diagonalization efficiency, we consider a diagonal operator
whose diagonal entries are sampled with repetition from the spectrum $\text{spec}(\D^*_{\rm min-max})$.
We sort them in an ascending order to facilitate comparisons with the min-max parametrization. We refer to this
parametrization as \textit{sampled min-max parametrization}.

If $\Hhat$ has a non-degenerate spectrum then instead of the min-max parametrization we can choose
 $\D^*_{\rm eigen}=\diag(\lambda_1, \lambda_2, ..., \lambda_n)$,
where $\lambda_1, \lambda_2,...,\lambda_n$ are eigenvalues of $\Hhat$ and we take
them in ascending order $\lambda_1\le\lambda_2\le...\le\lambda_n$.
Checking this operator should reveal whether the linear spectrum in min-max should instead have a non-linearity.

\subsection{Adaptive parametrizations}
We now explore some \textit{adaptive strategies}, in the sense that the parameters of
the diagonal operators are from now on considered to be optimized at each recursion step.

The first strategy is the \textit{dephasing channel parametrization} and is based on the fact the GWW DBI involves the dephasing channel
which restricts matrices to their diagonal. Thus, for each $k$, we set $\D_k=\Delta(\Hhat_k)$.
This strategy can also be used to initialize the BHMM DBI and inform it about the initial diagonal structure of the input matrix.
For example, $\Delta(\Hhat_{\text{TFIM}})= \sum _{j=1}^L  \Z_j$,
which coincides with a constant magnetic field. The XXZ model has $\Delta(\Hhat_\text{XXZ})= \delta\sum _{j=1}^L  \Z_j \Z_{j + 1}$, which means that the first step
of the GWW BHMM DBI agrees with a DBI parametrized by the classical nearest-neighbor (NN) Ising model.

Motivated by the above, we introduce the \textit{magnetic field parametrization},
which considers a local magnetic field in the $z$-direction
 \begin{align}
   \D_k (B^{(k)}) = \sum_{j=1}^L\alpha^{(k)}_j\Z_j\ ,
\end{align}
where $\alpha_j^{(k)}$ are the values of an artificial magnetic field of qubit $j$ in step $k$.
While $\alpha_j^{(k)}$ can be arbitrary, e.g. arising from gradient descent optimization, for BHMM DBI we show the cases where
$\alpha_j=1$ (constant), $\alpha_j=j$ (linear), and $\alpha_j=j^2$ (quadratic).
For adaptive DBI we optimize these using gradient descent in Qibo.
 These choices are motivated by the above min-max parametrization in the sense that by changing the $\alpha_j$ values we may try to
 lift degeneracies.

Next, we consider an ansatz which can have an entangling action on the qubits and
such extension of the non-interacting magnetic field can be useful for quantum many-body Hamiltonians.
More specifically we consider the classical Ising model with nearest-neighbor (NN) interactions
\begin{align}
    \D_k(B^{(k)},J^{(k)})=\sum_{j=1}^L\bigl(\alpha_j^{(k)}\Z_j + \beta_j^{(k)}\Z_{j+1}\Z_j\bigr),
\end{align}
which gives also the name to the ansatz.
For BHMM DBI we will choose uniform $\beta_i=1$ and for adaptive DBI again use gradient descent.

\begin{figure}[ht]
    \centering
    \includegraphics[width=0.9\textwidth, trim = 0 3cm 3cm 0, clip]{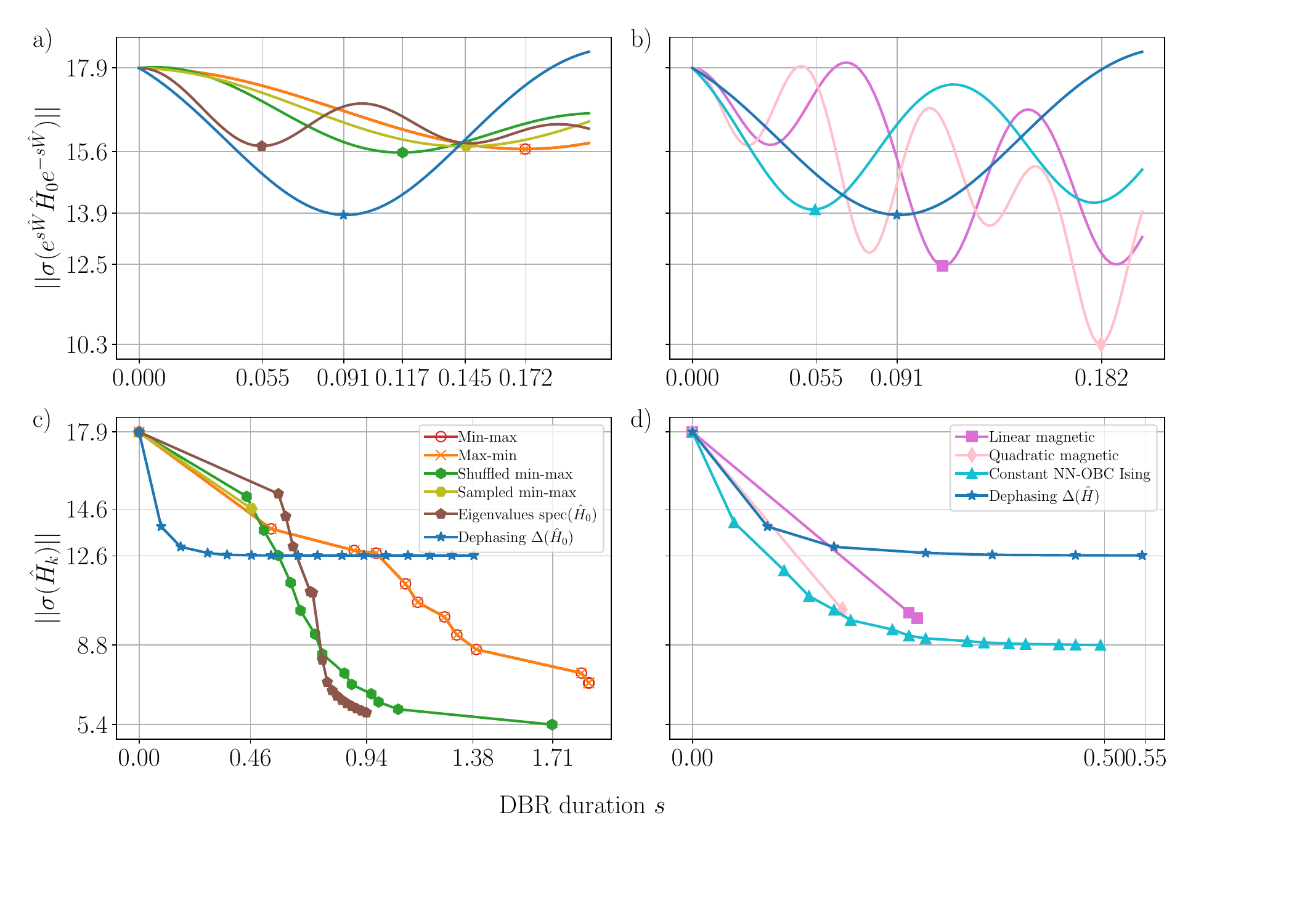}
    \caption{
DBRs (top) and DBIs (bottom) for the XXZ model with $L=5$ qubits and $\Delta = 0.5$.
In panel a) we show DBRs for the analytically motivated ansatze.
The GWW parametrization $\Delta(\h_0)$ yields the largest $\sigma$-decrease.
Bullets indicate the first local minima extracted numerically.
In panel b) we show DBRs for parametrizations which can be easily compiled using CNOT and single-qubit rotation gates.
Here the global minimum is not necessarily the first.
The GWW curve is repeated because it coincides with the NN Ising strategy.
In panels c) and d) we show the respective BHMM iteration. We stop, for each
adopted strategy, the DBI steps when they would have provided only marginal
gain at the expense of long evolutions.
Due to the symmetries of XXZ the GWW and magnetic field strategies yield block-diagonalization
as opposed to full diagonalization.
In panel d) for short $s$
the local $\D$ parametrization strategies perform comparably well with those in panel c).
These results set the benchmark when moving from analytically prescribed BHMM DBIs to adaptive DBIs.
}
    \label{fig:BHMM}
\end{figure}

Eigenstates of quantum many-body systems in the middle of the spectrum are very
strongly entangled so we also consider the case where there are local, i.e. 2-qubit, interactions
as above but the range of interaction is arbitrary
\begin{align}
    \D_k(B^{(k)},J^{(k)})=\sum_{j, j'=1}^L\bigl(\alpha_j^{(k)}\Z_j +
    \beta_{j,j'}^{(k)}\Z_{j}\Z_{j'}\bigr)\ .
\end{align}
We call this strategy \textit{all-to-all classical Ising model} and its all-to-all
interactions are a hallmark advantage of trapped-ion platforms.
To be specific we will set $\beta_{i,j}=1$ to be the uniform all-to-all parametrization.
Note that selecting $\beta_{i,j}=\delta_{i,j-1}$ reduces this model to the NN model.

Finally, we can explore what changes if we take more qubits to participate in an interaction.
Here we shall choose a binary string $\mu_k\in \{0,1\}^{L}$ and consider
\begin{align}
    \D_k = (Z_1)^{\mu_1}(Z_2)^{\mu_2}\ldots (Z_L)^{\mu_L}.
\end{align}
This allows any subset of qubits to interact.
The optimization in this case varies over the Hamming cube aided by solving the scheduling task.
We refer to this final strategy as \textit{Pauli-$Z$ product parametrization.}

All above models can be viewed as parametrizations for optimization over the parameters, e.g.
$\alpha \in  R^{L}$ and $\beta \in R^{L^2}$ for the all-to-all parametrization.
The extreme case is to allow all matrix elements to vary
\begin{align}
    \D_k=\sum_{i=1}^{2^L} d_i^{(k)}|i\rangle\langle i|
\end{align}

The disadvantage of this ansatz lies in its scalability as the number of parameters grows exponentially with the system size $\mathcal{O}(2^L)$.
\subsection{Benefits of normalized adaptive parametrizations}
Given our structuring of optimization strategies, varying the norm of the diagonal
operator $\hat{D}_k$ will not facilitate cost function gains.
To see this, let us consider a rescaling factor $r$ and we find that
if $\hat R = e^{-s_k [\hat D_k,\hat H_k]}$ was the optimal DBR for $\hat D_k$ then
$s_k' = s_k/r$ is the optimal duration for $\hat D_k' = r \hat D_k$.
This is because
$\hat R_k' = e^{-s_k'\ [\hat D_k'\,\hat H_k]}=\hat R_k$ is optimal
 whenever $\hat R_k$ is optimal.

We can use this fact together with the fact that adding a multiple of the identity
$zI$ to the diagonal operator leaves the corresponding DBR invariant, namely because
 $[\hat D_k +zI,\hat H_k] = [\hat D_k,\hat H_k] $.
 If we optimize the entries of $\hat{D}$ over  any interval $[a,b]$, then the optimal
 diagonal operator can be found.
 Indeed, if the true minimizer lies in the interval $[a',b']$ then the rescaling and shifting invariance
  of DBRs  allow to find $r$ and $z$ such that $[ra+z,rb+z]\subset [a',b']$.

  Finally, we point out that one may use projected gradient approaches to
optimize over unit-norm operators.
This can be implemented by taking a gradient descent step and dividing by the norm
after an update. Focusing on unit-norm parametrizations of diagonal operators is particularly
useful for the least-squares loss function $f_2$.
Here ordinary gradient descent will tend to focus on reducing the norm of $\hat{D}$
 instead of the $\Tr(\hat{H}_k \hat{D}_k)$ which is more imporant for diagonalization.
 In fact, when using projected gradient descent one could choose
 $f_2(\hat H_k)=\Tr(\hat{H}_k \hat{D}_k)$.
\subsection{Numerical results for the BHMM DBI}
In Fig.~\ref{fig:BHMM} we show the cost function landscape for the first DBI step which showcases how different parametrizations yield different $\sigma$-decrease.
We then show the performance of the BHMM DBI with the numerically obtained scheduling.
In Fig.~\ref{fig:BHMM} a) and c) we consider $\D$'s which are analytically motivated.
We find, that even though they possess analytical convergence guarantees it is slow for XXZ.
Additionally, we find that for the first few steps, degeneracies do not inhibit $\sigma$-decrease.
The DBIs in panel c) cannot be efficiently implemented on a quantum computer e.g.
eigenvalues have to be computed; due to their large number, their compilation could be difficult
even if they were known.
In Fig.~\ref{fig:BHMM} b) and d) we consider $\D$'s relevant for near-term quantum computers.
For the first few steps, these parametrizations perform similarly to those Fig.~\ref{fig:BHMM} a) and c), again despite having spectral degeneracies.

For all parametrizations, we find that BHMM DBIs lose efficiency and would need many steps with very short duration.
This motivates us to optimize $\D_k$ in every step and consider adaptive DBIs.

\subsection{Numerical results for the adaptive DBIs}
    \begin{figure}[ht]
        \centering
        \includegraphics[width=0.7\textwidth,trim = 0 0 0 1.5cm,clip]{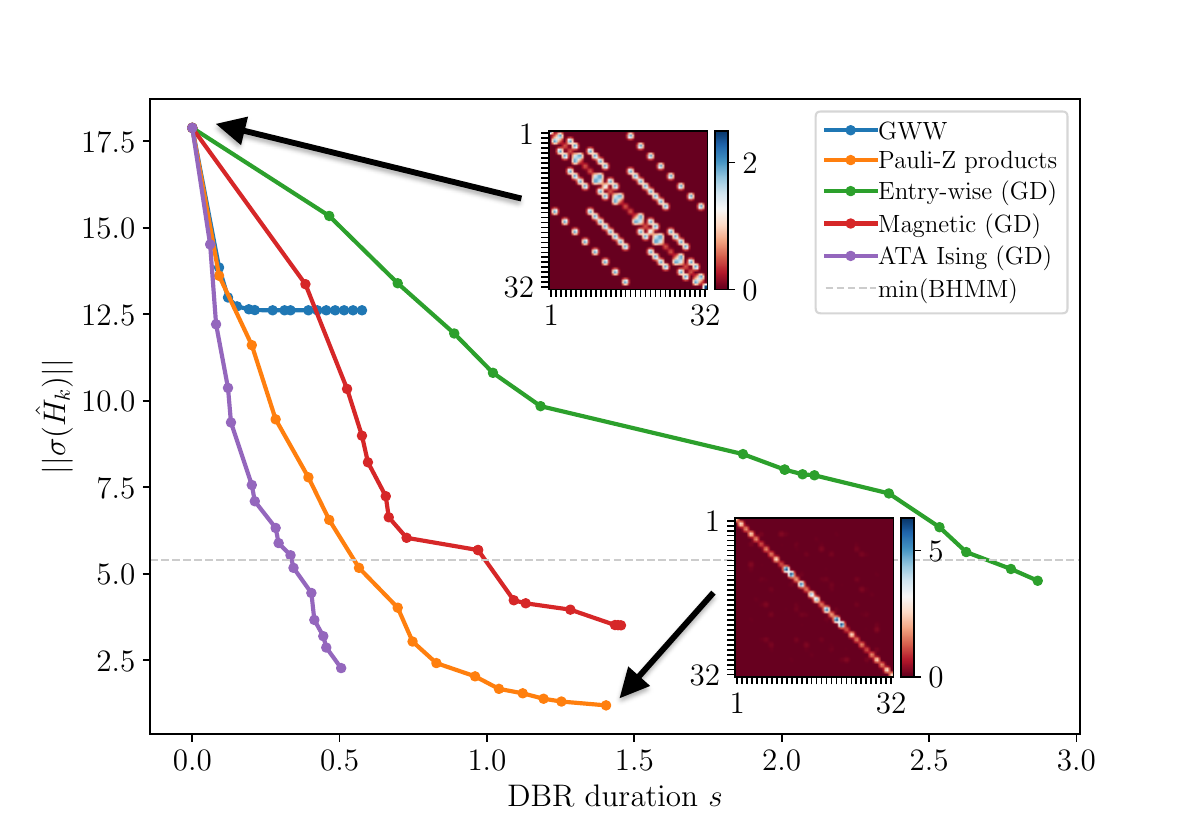}
        \caption{
            DBI applied to XXZ model with $L=5$ qubits, $\delta=0.5$ using variational strategies.
            Variational strategies, except for the GWW strategy (blue) surpass the best diagonalization achieved with BHMM (grey dotted line). See Fig.\ref{fig:BHMM}.
            Pauli-Z strategy achieved the most diagonalization but is difficult to implement physically.
            We also see that despite being more computationally exhaustive, gradient descent with computational ansatz (green) is not necessarily better than gradient descent with magnetic field (red) and Ising model ansatz (purple).
        }
        \label{fig:variational}
    \end{figure}

Fig.~\ref{fig:variational} shows results for adaptive DBIs again for the XXZ model on $L=5$ sites.
GWW DBI is both analytically motivated and adaptive but saturates due to XXZ symmetries.
Instead, we use standard gradient descent with an optimized learning rate for the magnetic field parametrization
$B$ and NN Ising parametrization $B$ and $J$ with open boundary conditions.
Gradient descent targets local minima that depend on initialization so choosing suitable parametrizations as above is important.

We find that within the same number of DBI steps the final cost function is significantly lower than for BHMM.
Thus if rounds of optimization are possible, which is true for future quantum hardware, adaptive DBIs should be considered.

\section{Group commutator}
\label{sec:gc}

A DBI is a sequence of DBRs.
However, DBR rotations cannot be directly implemented on a quantum computer.
Instead, quantum compiling in primitive gates should be used.
We compile approximations to DBR unitaries by means of group commutator formulas.
These consist of unitaries which then are simple exponentials of Hamiltonians and
so Hamiltonian simulations can be used which combine primitive gates into evolution operators $e^{-it\A}$, where $\A$ is a local Hamiltonian.

Specifically the DBR unitary $	\hat R_0 = e^{-s_0 [\D_0,\J_0]}$
can be approximated by the group commutator unitary
\begin{align}
	\v_0^{\text{(GC)}} = e^{i \sqrt{s_0}\J_0}e^{-i\sqrt{s_0}\D_0}e^{-i \sqrt{s_0}\J_0}e^{i\sqrt{s_0}\D_0}  = \hat R_0 +O(s_0^{3/2})\ .
\end{align}
Following Ref.~\cite{3rdOrderGC} we use the higher-order product formula (HOPF) with $\phi = \frac  12 (\sqrt 5 -1)$
\begin{align}
	\hat Q^{\text{(HOPF)}}_0 =
	e^{ i \phi\sqrt{s_0}\J_0}
	e^{-i \phi\sqrt{s_0}\D_0}
	e^{-i \sqrt{s_0}\J_0}
	e^{i(\phi+1)\sqrt{s_0}\D_0}
	e^{i (1-\phi) \sqrt{s_0}\J_0}
	e^{-i\sqrt{s_0}\D_0} = \hat R_0 +O(s_0^{2})\ .
\end{align}
One may note that when using the operators, the normalization results are no longer usable and so the optimization domain may correspond to the entire space.

\begin{figure}
	\centering
	\includegraphics[width=0.9\textwidth, trim = 0 0 0 0.3cm, clip]{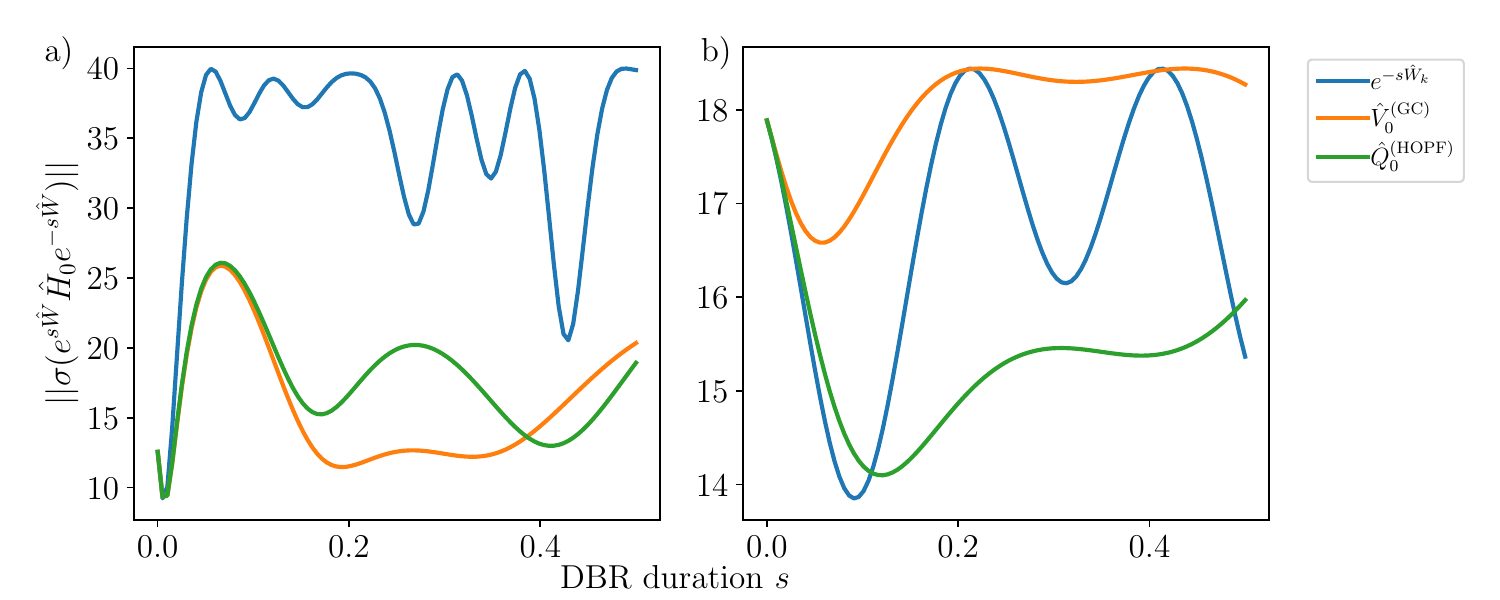}
	\caption{Group commutator approximation to DBR.
	In panel a) we apply DBR to the TFIM model with $L=5$ qubits, $h=3$ and see that both GC and HOPF are good approximations to the desired evolution up to $s<0.02$ and found the global minimum.
	Panel b) shows that on the XXZ model with $L=5$ qubits, $\delta=0.5$, HOPF performs considerably better than GC, which is far from the true minimum.
	}
	\label{fig:group_commutator}
\end{figure}

In Fig.~\ref{fig:group_commutator} we show the comparison of these approximations to the DBR unitary.
We found that consistently HOPF accurately matches the performance of DBR, even if the scheduling must be adjusted.
GC uses one less query to Hamiltonian simulation for $\J_0$ but it yields much less $\sigma$-decrease.

\section{Conclusions}
In this work, we presented strategies for optimizing double-bracket quantum algorithms.
The overall goal is to prepare eigenstate on a quantum computer and we discussed different diagonalization
cost functions which allow us to quantify the progress for this task.
We then described the formulation of DBQAs as being rooted in selecting the duration of the recursive steps
and selecting the DBR generators.

Fig.~\ref{fig:polynomial_fit} reports that the utilities featured in Qibo allow to reliably find schedulings that
yield effective diagonalization iterations.
We next discussed parametrizations of DBR generators aimed at obtaining effective DBIs through optimization.
We also showcased parametrizations suitable for BHMM DBIs which have analytical convergence guarantees.
Fig.~\ref{fig:BHMM} explored the role played by spectral degeneracies of the DBR generators.
For practical applications, we did not find it to be particularly key, especially when only a few recursion steps are available.
Additionally, we showed that BHMM DBIs can get stuck for exactly solvable models such as TFIM and XXZ  which motivates using adaptive DBIs.

Fig.~\ref{fig:variational} shows that adaptive DBIs achieve more diagonalization in fewer steps than the BHMM DBIs that we considered.
We further showed that numerical optimizations of the DBR generators within prescribed parameter spaces allow us to
extend the diagonalization convergence beyond what BHMM DBIs.
Finally, we discuss the strategies for approximating DBIs in a way that is viable for quantum compiling.
This is done by means of product formulas, e.g. the group commutator and Fig.~\ref{fig:group_commutator} shows that much better results are obtained by a higher-order product formula approximation to each DBR in a DBI.

In summary, we conceptually broke down the task of optimizing DBQAs into three distinct stages.
Our implementation in Qibo allowed us to showcase the performance of numerical optimization in each of these steps.
By this, double-bracket quantum algorithms are provided by a more systematic optimization methodology which we hope will aid
further progress in quantum compiling tasks such as eigenstate preparation of quantum many-body models involving large numbers of qubits.

\section{Acknowledgements}
The authors thank the DBI collaboration for discussions and the Qibo team for
support in developing the DBI code.
XL and MG are supported by the Presidential Postdoctoral
Fellowship of the Nanyang Technological University, Singapore. MR is supported by
CERN Doctoral Program through the CERN Quantum
Technology Initiative.


\end{document}